# Design and R&D of very forward calorimeters for detectors at future $e^+e^-$ collider

**Ivanka Bozovic-Jelisavcic [on behalf of the FCAL Collaboration][1]**
*Vinca Institute of Nuclear Sciences, University of Belgrade*
*M. Petrovica Alasa 12-14 Belgrade, Serbia*
*E-mail:* `ibozovic@vinca.rs`

Detectors at future $e^+e^-$ collider need special calorimeters in the very forward region for a fast estimate and precise measurement of the luminosity, to improve the hermeticity and mask the central tracking detectors from backscattered particles. Design optimized for the ILC collider using Monte Carlo simulations is presented. Sensor prototypes have been produced and dedicated FE ASICs have been developed and tested. For the first time, sensors have been connected to the front-end and ADC ASICs and tested in an electron beam. Results on the performance are discussed.

---

[1] Speaker

## 1. Introduction

In this paper we briefly summarize results of the simulation studies on the expected precision of the luminosity measurement and on the electron detection efficiency in the very forward region of a detector at a future linear collider. In these studies the ILD model [1] for ILC is assumed at a center-of-mass energy of 500 GeV. Including detailed study of systematic uncertainties, the feasibilty of the luminosity measurement at ILC with a precision of $10^{-3}$ has been revealed. Design studies are extended to the very forward region of a detector at CLIC at 1 TeV and 3 TeV center-of-mass energies. This has been elaborated as a part of the CLIC detector CDR [2].

In complement to the simulation studies, sensor prototypes and dedicated FE ASICs have been developed and tested in 4.5 GeV electron test beam at DESY for the two forward region callorimeters at ILC: LumiCal and BeamCal. These studies are ongoing and will be used to contribute to the ILC detector EDR.

## 2. Forward region layout

Two special calorimeters are foreseen for the instrumentation of the very forward region down to approximately 5 mrad, a luminometer designed to measure the rate of low angle Bhabha scatering (LumiCal) and BeamCal that will allow fast luminosity estimate and measurement of the beam parameters. Finely segmented and very compact calorimeters are needed to mach the requirements on precison. In addition, sensors of BeamCal have to be radiation hard since large ammount of beamstrahlung pairs (~10TeV/BX) will be deposited in this detector. Both detectors are realized as sampling calorimeters using tungsten as absorber. Silicon sensors are employed for LumiCal, while several technologies are an option for radiation-hard BeamCal sensors. Due to the high occupancies fast front-end electronics is needed.

## 3. Simulation studies

Simulation studies were mainly targeted to the expected precision of the luminosity measurement at ILC and on the efficient high-energy electron identification down to very low polar angles using two finely segmented and compact sampling calorimeters, LumiCal and BeamCal.

### 3.1 Luminosity measurement

Integrated luminosity measurement at ILC will be performed by counting Bhabha events reconstructed in the LumiCal. However, these counts have to be corrected for several systematic effects, the leading ones originating from 4-fermion processes as a physics background and the effective suppression of the Bhabha cross-section (BHSE) induced by beam-beam interaction effects. It has been shown [3] that event selection can be optimized to comprise the leading systematic effects by requiring the energy of Bhabha event to be more than 80% of the center-of-mass energy and by the proper definition of the effective counting volume (Figure 2).

In Table 1 two main sources of systematic uncertainty are considered at 500 GeV center-of-mass energy. All other souces contribute to the relative uncertainty of luminosity up to $1\cdot 10^{-3}$ [4]. Obtained results match the expectation from the physics program to measure luminosity with a precision at a permille level.

| BHSE | 4-f background/signal | Total contribution to the relative uncertainty of luminosity |
|---|---|---|
| (0.0±0.2)% | (1.6±0.4)·$10^{-3}$ | 2.0·$10^{-3}$ |

**Table 1:** BHSE and physics background to signal ratio in LumiCal. Uncertainties of the effects are estimated assuming 40% uncertainty for the 4-fermion cross-section from the NLO corrections and the bunch-width control of 5% for BHSE.

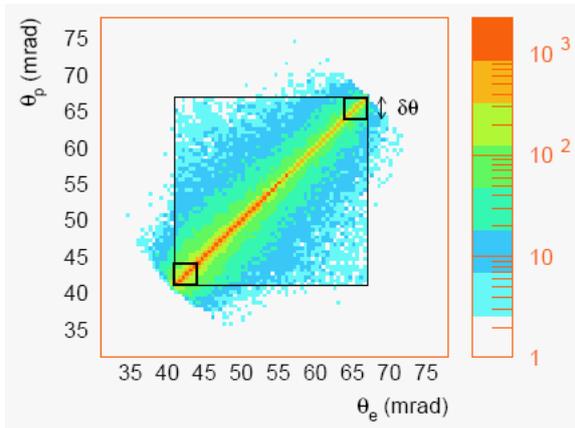
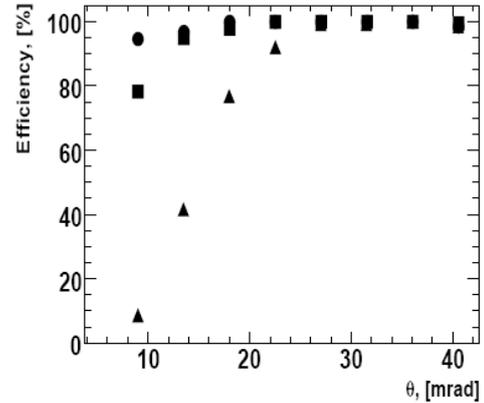

**Figure 2:** Definition of the effective Bhabha counting volume. Thin box stands for the detector fiducial volume. Thick boxes of the size δθ are excluded.

**Figure 3:** Electron identification efficiency in BeamCal on top of the beamstrahlung background, for 250 GeV (circles), 150 GeV (squares) and 75 GeV (triangles) electrons.

### 3.2 Electron identification efficiency

After each bunch crossing BeamCal will be hit by a large amount of beamstrahlung pairs depositing about 150 GeV of energy per buch crossing [4]. On the other hand, it is important for the new physics missing-energy signatures to be able to detect a single high-energy electron on top of the wide spread of beamstrahlung background. By performing an appropriate subtraction of the pair deposites and a shower-finding algorithm which takes into account the longitudinal shower profile, electron identification efficiencies are obtained as in Figure 3 [4]. This allows 4-fermion SM backround to be suppressed in a large fraction of the parameter space in e.g. in a SUSY searches [4].

## 4. Test-beam results

For the first time, silicon sensors developed for LumiCal and GaAs sensors for BeamCal have been connected to the front-end and ADC ASICs and tested in a 4.5 GeV electron beam at DESY. Set of measurements confirmed proper operation of the fully assembled detector module. Only a few ilustrative results are given in Figure 4 and 5 [3]. In Figure 4 the charge signal spectrum is shown, fitting nicely to a convolution of a Landau distribution and a Gaussian. Signal-to-noise ratio is estimated between 20 and 25, using the most probable value and the width of the pedestal. Similar results are obtained for BeamCal [3]. High statistics runs allowed a scan over 200 μm gap in between BeamCal pads. As shown in Figure 5, the signal size drops no more than 10% in the gap region. Again, similar results are obtained by scaning over the 100μm LumiCal gaps [3].

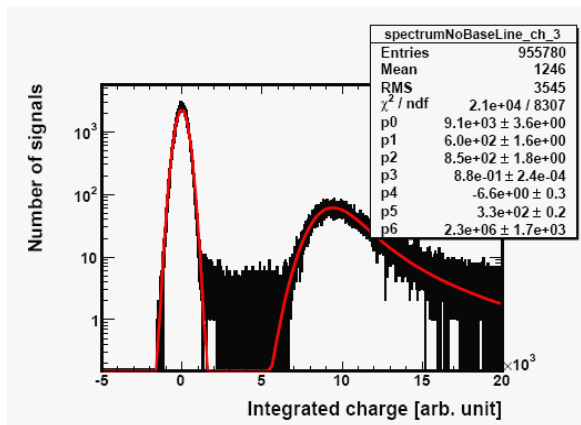

**Figure 4:** LumiCal: Charge signal spectrum.

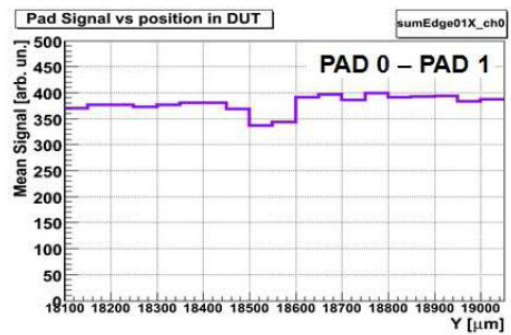

**Figure 5:** BeamCal: Mean signal distributon vs. particle impact point on the sensor. Drop over gap between pads is visible.

## 5. Conclusions

Instrumentation of the very forward region of a detector at ILC has been designed. Parameters relevant for the physics program have been estimated from simulation and found to match the requirements. Prototypes of major components such as sensors, front-end and ADC ASICs have been developed, produced and tested as a whole. Measured performance fulfils the specification derived from simulation.